\documentstyle[11pt,epsfig]{article}

\title{\bf A New interpretation of MOND based on Mach principle and an Unruh
like effect}
\author{F. Darabi\thanks{email: f.darabi@azaruniv.edu; Fax:+98-412-4327541 }
\\ {\small  Department of Physics, Azarbaijan
University of Tarbiat Moallem, Tabriz 53741-161, Iran}\\{\small
Research Institute for Astronomy and Astrophysics of Maragha
(RIAAM), Maragha 55134-441, Iran}}
\begin{document}
\maketitle

\begin{abstract}
A new interpretation is introduced for MOND based on the Sciama's
interpretation of Mach principle and an Unruh like effect, in the context
of a generalized equivalence principle. It is
argued that in a locally accelerated frame with acceleration $a$ the
appearance of a Rindler horizon may give rise to a constant
acceleration $a_0$ as the local properties of cosmological horizon
or Hubble length. The total gravitational acceleration inside this
frame becomes the combination of $a$ with $a_0$. For $a\gg a_0$, the
conventional gravitational mass $m_g$ interacts with the dominant
acceleration as $m_g a$ and application of Sciama's interpretation
leads to the standard Newtonian dynamics. For $a\ll a_0$, however, a
reduced gravitational mass $\bar{m}_g$ interacts with the dominant
acceleration as $\bar{m}_g a_0$ and the application of Sciama's
interpretation on this reduced gravitational mass leads to MOND.
This introduces a third proposal for MOND: {\it The modification of
gravitational mass}.
\end{abstract}
\vspace{2cm}
\newpage

\section{Introduction}
It is well known that classical Newtonian dynamics fails on galactic
scales. There is astronomical and cosmological evidence for a
discrepancy between the dynamically measured mass-to-light ratio of
any system and the minimum mass-to-light ratios that are compatible
with our understanding of stars, of galaxies, of groups and clusters
of galaxies, and of superclusters. It turns out that on large scales
most astronomical systems have much larger mass-to-light ratios than
the central parts. Observations on the rotation curves have turn out
that galaxies are not rotating in the same manner as the Solar
System. If the orbits of the stars are governed solely by
gravitational force, it was expected that stars at the outer edge of
the disc would have a much lower orbital velocity than those near
the middle. In fact, by the Virial theorem the total kinetic energy
should be half the total gravitational binding energy of the
galaxies. Experimentally, however, the total kinetic energy is found
to be much greater than predicted by the Virial theorem. Galactic
rotation curves, which illustrate the velocity of rotation versus
the distance from the galactic center, cannot be explained by only
the visible matter. This suggests that either a large portion of the
mass of galaxies was contained in the relatively dark galactic halo
or Newtonian dynamics does not apply universally.

The dark matter proposal is mostly referred to Zwicky \cite{Zwicky}
who gave the first empirical evidence for the existence of the unknown
type of matter that takes part in the galactic scale only by its
gravitational action. He found that the motion of the galaxies of
the clusters induced by the gravitational field of the cluster can
only be explained by the assumption of dark matter in addition to
the matter of the sum of the observed galaxies. Later, It was
demonstrated that dark matter is not only an exotic property of
clusters but can also be found in single galaxies to explain their
flat rotation curves.

The second proposal results in the modified Newtonian dynamics
(MOND), proposed by Milgrom.  MOND, as a phenomenological theory,
may be interpreted as 1) a modification of inertia, 2) a
modification of gravity \cite{MOND}. The well known second law of motion states that an
object of mass $m$ subject to a force $F$ undergoes an
acceleration $a$ by the simple equation $F=ma$. However, it has
never been verified for extremely small accelerations which are
happening at the scale of galaxies. Based on the modification of
inertia the proposition made by Milgrom was the following
\begin{equation}
F=m\mu(\frac{a}{a_0})a,\label{1}
\end{equation}
\begin{equation}
\mu(x)=\left \{ \begin{array}{ll} 1 \:\: \mbox{if}\:\: x\gg 1
\\
x \:\: \mbox{if}\:\: x\ll 1,
\end{array}\right.\label{2}
\end{equation}
where $a_0=1.2\times10^{-10} ms^{-2}$ is a proposed new constant.
The acceleration $a$ is usually much greater than $a_0$ for all
physical effects in everyday life, therefore $\mu(a/a_0)$=1 and
$F=ma$ as usual. However, at the galactic scale where $a \ll a_0$
we have the modified dynamics $F=m(\frac{a^2}{a_0})$ leading to a
constant velocity of stars on a circular orbit far from the center
of galaxies.

There are possible approaches to MOND inertia: deriving effective
inertia from interactions with the medium such as vacuum, seeking
a new symmetry like Lorentz symmetry that forces a form of the
free actions compatible with MOND, and using Mach principle as a
new connection between the universe at large and local inertia
\cite{MOND}. However, to the authors knowledge, no attention has
been paid to study the MOND in the framework of Sciama's
interpretation of Mach principle in that inertial forces are
interpreted as interaction forces containing an
acceleration-dependent term \cite{Sciama}. Based on this interpretation,
one may find the inertia of an object as the gravitational interaction
of its gravitational mass with the distant matter distribution in the universe.
So, if according to Milgrom's idea the inertia of an object is to change
at very small acceleration, it is reasonable to interpret this modified inertia
based on the same Sciama's interpretation.

The main observation in the interpretation of MOND is the connection
of $a_0$ with the cosmic features hinting that MOND might result
only in the context of a non-Minkowskian universe, with $a_0$
reflecting the departure from flatness of space time \cite{MOND}.

In this paper, we do not favor MOND to dark matter models and vice versa
or do not claim that Sciama's ansatz is correct. However, if MOND has any
relation to reality and provided that Sciama's ansatz is reasonable enough
to describe the laws of motion, we would like to revisit the MOND from Sciama's interpretation of Mach principle.

\section{Sciama's interpretation of Mach principle }

In this section, we partly use of Ref.\cite{Gogberashvili} to establish the
Sciama's interpretation of Mach principle.
According to Mach, in any two-body interaction the influence of all
other matter inside their causal sphere should be taken into account
\cite{Mach}. For instance, in the process of gravitational
interaction of close objects one can replace the distant universe by
a spherical shell of the effective mass $M$ and the effective radius
$R$. This shell may act as a gravitational Faraday cage inside of
which a constant gravitational potential exists as
\begin{equation}
\phi=-\frac{GM}{R}.\label{5}
\end{equation}
For an inertial particle the universal field $\vec{E}$ is zero
\begin{equation}
\vec{E}=-\vec{\nabla} \phi=0.\label{6}
\end{equation}
For accelerated particles, however, similar to the induction law in
electrodynamics we have \cite{Sciama}
\begin{equation}
\vec{E}=-\vec{\nabla} \phi-\frac{\phi}{c^2}\vec{a}=-\frac{\phi}{c^2}\vec{a},\label{7}
\end{equation}
where $\vec{a}$ is the acceleration vector and $c$ is the light velocity.
Considering a homogeneous and isotropic distribution of matter in
the universe with the average mass density $\rho$ one may write
\begin{equation}
-\frac{\phi}{c^2}=\frac{G\rho V}{Rc^2}=\frac{2\pi G \rho
R^2}{c^2}=\frac{2\pi G \rho}{H^2},\label{8}
\end{equation}
where the volume $V$ of the universe is considered as a sphere with
the Hubble radius
\begin{equation}
R=\frac{c}{H}.\label{9}
\end{equation}
Introducing the critical mass density
\begin{equation}
\rho_c=\frac{3H^2}{8\pi G},\label{10}
\end{equation}
and using Eq.(\ref{8}) the following relationship is obtained with a reasonable
degree of precision
\begin{equation}
\phi\approx-c^2.\label{11}
\end{equation}
It is to be noted that this relationship is exact at the Planck time
\begin{equation}
\phi=\frac{GM_p}{L_p}=-c^2,
\end{equation}
where $M_p$ and $L_p$ are the Planck mass and length, respectively.
Therefore, we assume that the condition (\ref{11}) would be valid in its
exact form $\phi=-c^2$ and hence the gravitational potential would remain unchanged for all the history of the expanding universe \cite{Kenyon}.

The exertion of the gravitational field of
the whole universe (\ref{7}) on an accelerated particle with the
gravitational mass $m_g$ leads to the standard
expression of the inertial force
\begin{equation}
\vec{F}=m_g\vec{E}=m_i\vec{a},\label{12}
\end{equation}
where the concept of inertial mass $m_i$, as Mach would desire, is
appeared as the measure of the gravitational interaction of the
particle with the whole universe as \footnote{This relation written as $m_i
c^2+m_g \phi$ implies that the total energy (rest and gravitational) of a
particle in the universe is zero, a fact which is certified by theory of
inflation in that the total energy (Hamiltonian) of the universe is zero.}
\begin{equation}
m_i=-m_g\frac{\phi}{c^2},\label{13}
\end{equation}
which using of $\phi=-c^2$ implies for the equality $m_g=m_i$ as the strong
equivalence principle.

\section{A universal constant acceleration}

The existence of a universal acceleration $\vec{a_0}$ in the Milgrom's
model of dynamics may reveal interesting relation between MOND and
Mach Principle in that MOND can represent a feature of the universe as a
whole on local dynamics. It is well known that $\vec{a_0}$ may be expressed in
terms of some cosmological quantities.
There are, in fact, some quantities with the dimensions of an acceleration,
that can be constructed from cosmological parameters as \cite{MOND}
\begin{equation}
a_{ex} \equiv cH_0,\label{14}
\end{equation}
\begin{equation}
a_{cu}\equiv \frac{c^2}{R_c},\label{15}
\end{equation}
or
\begin{equation}
a_{\lambda}\equiv {c^2}|\lambda|^{\frac{1}{2}},\label{16}
\end{equation}
where $H_0$ is the present expansion rate of the universe (the Hubble constant),
, $R_c$ is the radius of spatial curvature of the universe and $\lambda$
is the cosmological constant.
The problem of identification of $\vec{a_0}$ with one of the above candidates would be an important step toward constructing an underlying theory for MOND.
The important point is that if $\vec{a_0}$ as a cosmological quantity do physically exist, then it should become relevant for the particle
dynamics once the particle is coupled with the cosmos, as a whole.

From equivalence principle of general relativity we know there is a deep relation between acceleration and gravitation. A frame linearly accelerated relative to an inertial frame in the Minkowskian background of {\it special relativity} is locally identical to a frame at rest in a gravitational field. It is then obvious that in order to accommodate a universal constant acceleration $\vec{a_0}$ within this equivalence principle, one may generalize this statement in such a way that large enough above this critical acceleration we get the Newtonian dynamics and small enough below $\vec{a_0}$ we obtain MOND.

We first suppose a frame which is linearly accelerated by an external agent
relative to an inertial frame with acceleration $\vec{a}$. The equivalence principle states that as long
as the gravitational field in this frame is produced by the local acceleration
of this frame with respect to an external inertial frame, the inertial and gravitational
masses are indistinguishable. This is known as the strong equivalence principle
(SEP). One may wish to show this by resorting to Sciama's interpretation of Mach principle: The observer inside the accelerated frame
will observe a massive object freely falling with acceleration $\vec{a}$. If, according
to Sciama's interpretation of Mach principle, this observer believes in inertia as a kind of interaction with the universe, he should write down this equality
according to Eqs.(\ref{7}), (\ref{12}) and using $\phi=-c^2$ as
\begin{equation}
m_g\vec{E}=m_g \vec{a}=m_i \vec{a},\label{17}
\end{equation}
where $m_i=m_g$, as the representation of (SEP), results due to the interpretation
of inertial force $m_i \vec{a}$ as a kind of gravitational interaction $m_g \vec{a}$.

Now, suppose there is a constant universal curvature in the
neighborhood of every small region in the space. So, if there is a
rest frame in this neighborhood, then the isotropy and homogeneity
of the universe implies that this constant curvature does not
produce a gravitational field with a local preferred direction
inside this rest frame. If however, this frame is accelerated by an
external agent in a given direction, according to equivalence
principle there will be a gravitational field inside the frame with
the same magnitude but opposite direction as $\vec{a}$. Moreover,
the isotropy inside this accelerated frame is broken down due to the
preferred  gravitational direction and it is plausible that this
symmetry breaking may stimulate an extra local gravitational field
originated by the constant universal curvature, with the same
direction as that of $\vec{a}$. Note that the isotropy is broken
down just for accelerated frames and not inertial ones with uniform
motions. This may be similar to the well known Unruh effect in that
a thermal effect appears merely in an accelerated frame which breaks
down the isotropy of the vacuum due to the appearance of an apparent
event horizon. The Rindler coordinate system or frame describes a
uniformly accelerated frame of reference in Minkowski space. It is
well known that the Rindler spacetime has a horizon and gives the
local properties of {\it black holes} and {\it cosmological
horizons}. We know the Unruh effect is the near-horizon form of the
Hawking radiation. In the same way, one may suppose a non trivial
gravitational effect appears in the accelerated frame as a local
properties of {\it cosmological horizon} or Hubble length $H_0$
which is related to the constant curvature of the universe through
the $R_c$, namely the radius of spatial curvature of the universe.

One may also consider the emergence of the nonzero universal acceleration $a_0$ as an Unruh like effect, within the framework of holographic principle. The holographic principle is a
property of quantum gravity which states that the description of a
volume of space can be considered as encoded on a boundary, like a
gravitational horizon, to that region. This theory suggests that the
entire universe can be seen as a two-dimensional information
structure "painted" on the cosmological horizon, namely the universe
may be like a gigantic hologram. In this regard, one may consider
the {\it local} acceleration $a$ as the motion with respect to the
neighborhood volume of space, and the {\it universal} acceleration
$a_0$ as the motion with respect to the entire boundary of the observable
universe, namely the cosmological horizon.

The appearance of this extra gravitational field which is related to
$H_0$ or $R_c$ may account for a constant universal acceleration
$a_0\approx a_{ex} $ or $a_0\approx a_{cu}$ inside the accelerated
frame, as Milgrom has proposed \cite{MOND}. Therefore, the observer
in this frame will observe an interaction of the gravitational mass
of the object with the total acceleration
\begin{equation}
\vec{a}_{T}=\vec{a}+\vec{a_0}.\label{17'}
\end{equation}
This gravitational interaction must be balanced by the inertial force $m_i\vec{a}$
where $m_i$ is the {\it constant} inertial mass which is coupled just to
$\vec{a}$ with inertial origin (produced by an external agent as mentioned
above) and has nothing to do with $\vec{a_0}$ which has a gravitational origin
as a local properties of cosmological horizon (produced by an Unruh like effect).

In fact, $\vec{a_0}$ makes an absolute demarkation between the
gravitational and inertial masses and they are no longer equivalent
inside this locally accelerated frame. Since this difference is
caused by an extra gravitational field $\vec{a_0}$ and the inertial
properties of an object do not change (the external accelerating
agent does no change) it is plausible to assume that in this frame
the inertial mass of the object is perfectly constant and it is the
behavior of gravitational mass which is changed due to the
appearance of $\vec{a_0}$.

For $a\gg a_0$ the dominant acceleration is $\vec{a}_{T}=\vec{a}$.
So, the gravitational interaction occurs between the gravitational
mass $m_g$ and $\vec{a}$ which is to be balanced by the inertial
force as
\begin{equation}
m_g\vec{a} =m_i \vec{a}.\label{18}
\end{equation}
This case was already discussed in the previous section and led to
the Newtonian dynamics. For $a\ll a_0$ the dominant acceleration is
$\vec{a}_{T}=\vec{a_0}$. In this case, the gravitational interaction
occurs between an effective gravitational mass $\bar{m}_g$ and
$\vec{a_0}$ in order to be balanced with the same (unchanged)
inertial force as
\begin{equation}
\bar{m}_g\vec{a_0} =m_i \vec{a},\label{18'}
\end{equation}
which leads to a {\it reduced} gravitational mass
\begin{equation}
\bar{m}_g =m_i \frac{a}{a_0}.\label{18''}
\end{equation}
The important and key point is that the reduced gravitational mass
$\bar{m}_g$ is interpreted as the {\it gravitational charge} against
the universal gravitational field $\vec{a_0}$ and all other
gravitational sources with cosmological origin. On the other hand,
the conventional gravitational mass $m_g$ is interpreted as the {\it
gravitational charge} against the local gravitational fields
$\vec{a}$ . Now, suppose our observer in the accelerated frame would
like to use again Sciama's interpretation of Mach principle to
evaluate the gravitational interaction of the object with the
cosmological potential $\phi$. To this end, using (\ref{18}) and
$\phi=-c^2$ he should write down as follows
\begin{equation}
\bar{m}_g\vec{E}=-m_i\frac{a}{a_0}\frac{\phi}{c^2}\vec{a}=m_i\frac{a}{a_0}\vec{a},\label{19}
\end{equation}
where, as mentioned above, the reduced gravitational mass
$\bar{m}_g$ as the {\it gravitational charge} against the
gravitational sources with cosmological origin interacts with the
gravitational field $\vec{E}$ which has a cosmological origin
$\phi$. The resultant gravitational interaction manifests as the law
of motion in the modified dynamics which Milgrom has proposed. The
observer in the accelerated frame may apply this inertial force for
the freely falling object which appears to be under a local real
central gravitational force $F\sim r^{-2}$ as
\begin{equation}
\frac{G m_g M}{r^2}=m_i\frac{a^2}{a_0}.\label{20}
\end{equation}
Note that in the L.H.S of force law of gravitation, the conventional
gravitational mass $m_g$ is appeared since according to our
assumption it is interacting with the {\it local} gravitational
force $F\sim r^{-2}$. Therefore, using (SEP) as $m_g=m_i$ the
observer arrives at
\begin{equation}
a=\frac{\sqrt{G M a_0}}{r}.\label{21}
\end{equation}
According to equivalence principle the results obtained in this accelerated
frame is the same for an observer at rest in a gravitational field. Therefore,
the central (freely falling) acceleration of an object for which $a\ll a_0$ at a distance $r$ from
the center of a galaxy is the same as (\ref{21}) which may justify the galaxy's rotation curve.

\newpage
\section{Conclusion}

In this paper, we have interpreted the MOND based on the Sciama's
interpretation of Mach principle, an Unruh like effect and the
equivalence principle. We have shown that the equivalence principle
may be generalized to incorporate the constant acceleration $a_0$
which is produced in an Unruh like effect. The price is paid by
introducing a reduced gravitational mass. At large accelerations $a
\gg a_0$, $a$ being the acceleration of the local frame, the
dominant interaction occurs between the {\it conventional}
gravitational mass and $a$ so that application of Sciama's ansatz
leads to Newtonian dynamics.  At small accelerations $a \ll a_0$,
the dominant interaction occurs between the {\it reduced}
gravitational mass and $a_0$ so that application of Sciama's ansatz
leads to Modified Newtonian dynamics. This approach introduces a
third proposal for MOND as {\it modification of gravitational mass}
beside the two previous proposals, namely {\it modifcation of
inertia} and {\it modifcation of gravity}.

\section*{Acknowledgment}

This work has been supported financially by Research Institute for
Astronomy and Astrophysics of Maragha (RIAAM).

\end{document}